\documentclass[fleqn, usenatbib]{mnras}

\usepackage{newtxtext,newtxmath}

\usepackage[T1]{fontenc}

\usepackage{graphicx}	
\usepackage{amsmath}	
\usepackage{mathrsfs}
\usepackage[flushleft]{threeparttable}
\usepackage{siunitx}
\usepackage{booktabs, caption}
\usepackage{float}
\usepackage{pdflscape}
\usepackage{ulem}
\usepackage{physics}

\newcommand{\GG}[1]{}

\title[PAH Kinematics from PCA]{The Kinematics of Polycyclic Aromatic Hydrocarbons (PAHs) in Galaxies revealed by Principal Component Analysis (PCA) tomography with JWST/NIRSpec.}

\author[F. R. Donnan et al.]{
Fergus R. Donnan, $^{1}$\thanks{E-mail: fergus.donnan@physics.ox.ac.uk}  Dimitra Rigopoulou,$^{1, 2}$  Ismael Garc{\'i}a-Bernete $^{1}$\\
$^{1}$Department of Physics, University of Oxford, Keble Road, Oxford, OX1 3RH, UK\\
$^{2}$School of Sciences, European University Cyprus, Diogenes street, Engomi, 1516 Nicosia, Cyprus\\
}

\date{Accepted XXX. Received YYY; in original form ZZZ}

\pubyear{2024}

\begin{document}
\label{firstpage}
\pagerange{\pageref{firstpage}--\pageref{lastpage}}
\maketitle

\begin{abstract}
Polycyclic Aromatic Hydrocarbons (PAHs) are organic molecules which comprise the smallest particles of dust in the interstellar medium (ISM). Due to their broad/complex emission profiles, obtaining kinematics is a challenge with traditional methods, especially before the advent of the JWST. In this work we employ Principal Component Analysis (PCA) tomography to analyse JWST/NIRSpec IFU data of 3 nearby luminous infrared galaxies (LIRGs), namely, NGC 3256 N, NGC 3256 S and NGC 7469. We detect the signature of rotation in the second principal component of the 3.3 $\mu$m PAH feature in all three targets. We construct velocity maps from the principal components for the 3.3 $\mu$m PAH feature, Br$\beta$ (2.625$\mu$m) and molecular hydrogen, H$_2$ 1-0 S(1) (2.12$\mu$m). We find that in each target, the PAHs qualitatively follow the rotation of the galaxy, consistent with the rotational signature derived from Br$\beta$ and H$_2$. There are however some differences, with the PAH rotation in NGC 3256 N appearing at a different position angle, which suggest differences in the motion of the dust as compared to the gas. This kind of analysis opens a new window into this key component of the ISM.
\end{abstract}




\section{Introduction}
Polycyclic Aromatic Hydrocarbons (PAHs) are molecules comprised of numerous carbon rings bonded with hydrogen and are the smallest particles of dust. PAH molecules are excited by UV photons from young stars \citep[e.g][]{Tielens2008} and therefore are thought to be excellent tracers of star-formation \citep[e.g.][]{Rigopoulou1999, Peeters2004, Shipley2016}. These molecules are extremely abundant in galaxies, both locally \citep[e.g.][]{Brandl2006, Smith2007} and in the early universe \citep[e.g.][]{Spilker2023}, playing an important role in the ISM \citep[e.g.][]{Bakes1994} and dominating the mid-IR spectra of galaxies. A missing diagnostic of this key component of the ISM are its kinematics.

Originating from the bending and stretching of C-H and C-C bonds, \citep[e.g.][]{Tielens2008, Li2020}, PAH features have broad and complex profiles, consisting of multiple emission components \citep[e.g.][]{Rigopoulou2021, Draine2001}. PAH profiles are significantly broader than any typical velocity signature in a galaxy, and thus measuring a velocity shift is challenging. Moreover, the intrinsic shape of a given PAH profile can vary based on the conditions of the ISM such as the hardness of the radiation field, the charge and the mass \citep[e.g.][]{Peeters2002, Candian2015, Shannon2019}. This intrinsic broad shape also means there is no well-defined line centre that one can measure deviations from. For these reasons, deriving kinematics from PAHs are extremely challenging to measure.

PAHs are a key component of the ISM and in particular in photo-dissociation regions (PDRs) \citep[e.g.][]{Hollenbach1997}, where PAHs are excited on the boundary between the ionised HII region and molecular gas \citep[e.g.][]{Andrews2015, Chown2023}. Whether the motion of PAHs follow the molecular gas or ionised gas from stars in unknown, and in systems with strong outflows, the motion of PAHs is particularly interesting. This is especially true considering outflows are known affect the PAH population, resulting in altered ratios between different PAH bands \citep[e.g.][\textcolor{blue}{Garc{\'i}a-Bernete submitted.}]{Beirão2015, Garcia-Bernete2022c}.




Following \citet{Gorski2023} and utilising the method outlined in \citet{Steiner2009}, we apply Principal Component Analysis (PCA) tomography to identify signatures of PAH kinematics from the data where traditional methods (such as modelling the profile with a Gaussian) are not suitable for the reasons outlined above.

PCA is a method of dimensionality reduction by transforming the data into a new coordinate system of uncorrelated principal components, each of which describe decreasing variance within the data. PCA has been widely used in astronomy, such as with Spitzer spectra \citep[e.g.][]{Wang2011, Hurley2012, Sidhu2022}, as it provides a blind technique that can extract information from higher dimensions in the data. When applied to data cubes which contain two spatial dimensions and a spectral dimension, the PCA reduction extracts eigenvectors (in this case eigenspectra) using the information contained in the spatial axes. This allows spectral signals that correlate with some spatial distribution, that may be hidden in the data, to be uncovered \citep[e.g.][]{Meier2005, Steiner2009, Ricci2011, Gorski2023}.

By isolating emission features in the data, kinematic signatures such as rotation or inflowing/outflowing material can be revealed, typically as the second or third principal component \citep[e.g.][]{Gorski2023}. 

In this letter we first summarise the technique of \citet{Steiner2009} in Section \ref{sec:PCA} before applying the technique to JWST/NIRSpec data of three targets \citep[See][for more details]{Donnan2024} in Section \ref{sec:PAHKinematics}. We apply PCA to the 3.3 $\mu$m PAH, Br$\beta$ (2.62 $\mu$m), H$_2$ (2.12 $\mu$m) and compare the kinematics of each.

\section{JWST/NIRSpec Data}
We use NIRSpec IFU data from Director's Discretionary Early Release Science Program 1328 (PI: Lee Armus \& Aaron Evans) of three galaxies, NGC 3256 N, NGC 3256 S and NGC 7469. We use the data presented in \citet{Donnan2024}, reduced with modifications to the default pipeline as described in \citet{Garcia-Bernete2023, Pereira-Santaella2023}.

NGC 3256 is a late-stage merger of two galaxies, where each nucleus was observed separately. The northern component is a near face-on spiral hosting a star-forming nucleus \citep[NGC 3256 N,][]{Sakamoto2014, Lira2008}, while the southern component is an edge-on spiral hosting an obscured AGN \citep[NGC 3256 S,][]{Ohyama2015, Pereira-Santaella2023, Donnan2024}. Both nuclei drive molecular outflows \citep{Sakamoto2014, Pereira-Santaella2023}.

NGC 7469 is a barred spiral galaxy with a type 1 AGN residing within a starburst ring \citep[e.g.][]{Osterbrock1993, Diaz-Santos2007, Garcia-Bernete2022b, Lai2022, Lai2023}. For a plot of the field-of-view of the NIRSpec observations for each target, see Fig. 6 of \citet{Donnan2024}.

\section{Principal Component Analysis tomography}
\label{sec:PCA}
\subsection{PCA}

Before applying the PCA technique, we first prepare the data cubes by isolating the emission feature within the cube and subtracting the continuum. We use a simple local continuum by fitting a straight line between either side of the emission feature. This produces a cube with the only the emission feature present. Additionally we subtract the mean flux at each wavelength to normalise the data cube, a standard practice for PCA. The following summarises the technique described in \citet{Steiner2009}.

The resulting data cube has a total of $m$ spectral pixels and $n=\mu \times \nu$ spatial pixels, where $\mu$ and $\nu$ are the number of pixels in $x$ and $y$ respectively. From this we additionally reorganise the data cube, $\mathbf{I}_{x, y, \lambda}$ by stacking the spatial dimensions, $x$ and $y$, into a single coordinate, $\beta$ where 
\begin{equation}
\label{eqn:CoordTrans}
    \beta = \mu (x-1) +y.
\end{equation}
The reorganised data cube, $\mathbf{I}_{\beta, \lambda}$ is transformed using 
\begin{equation}
    \mathbf{T}_{\beta, k} = \mathbf{I}_{\beta, \lambda} \cdot \mathbf{E}_{\lambda, k}
\end{equation}
where $\mathbf{T}_{\beta, k}$ is the transformed data which can be mapped back to the original spatial dimensions $\mathbf{T}_{x, y, k}$, using equation (\ref{eqn:CoordTrans}). These are the tomograms for each principal component, $k$, where $k=1$ gives the first principal component with subsequent components describing decreasing variance within the data. $\mathbf{E}_{\lambda, k}$ contains the eigenvectors of each component, $k$, which in this case we call eigenspectra, as these are functions of wavelength, $\lambda$. The tomogram, $T_{x, y, k}$, therefore describes the spatial correlation of its corresponding eigenspectrum and can thus be used to reveal signals hidden in the data that are spatially correlated.

\begin{figure}
	\includegraphics[width=\columnwidth]{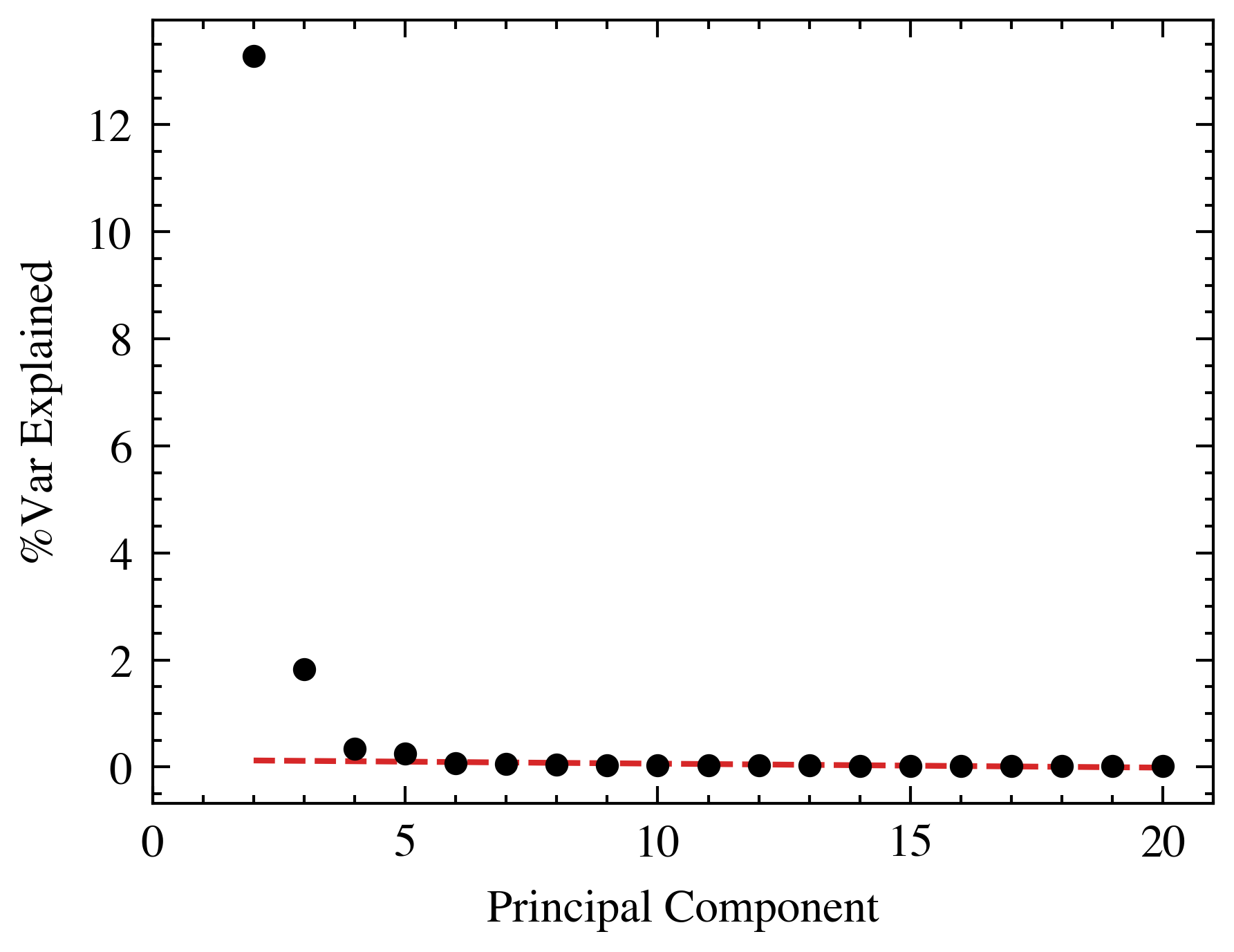}
    \caption{Scree test for the PCA applied to Br$\beta$ for NGC 3256 N, which shows the first three principal components are significant. The 1\textsuperscript{st} component explains 83.5$\%$ of the variance and appears of the scale. The red dashed line shows a linear fit the the higher order components where no useful information is contained. }
    \label{fig:Scree} 
\end{figure}

To establish if the components are useful, we can perform a Scree test \citep{Cattell1996}, which provides a measure of the variance explained by each principal component. This is shown in Fig. \ref{fig:Scree} for Br$\beta$ (2.62 $\mu$m) in NGC 3256 N and demonstrates that the first three components significantly above the point where the plot levels off.

\begin{figure*}
	\includegraphics[width=15cm]{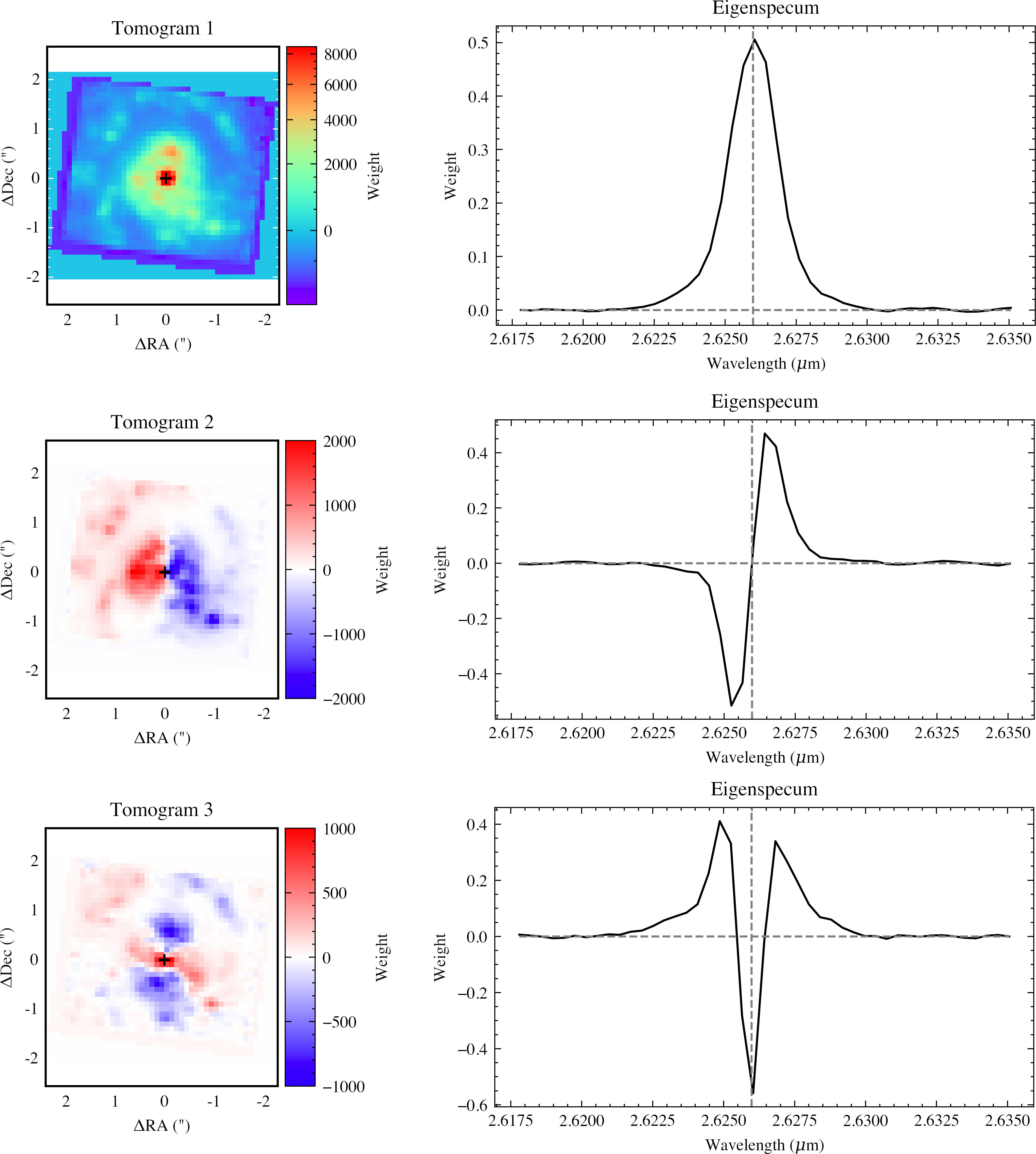}
    \caption{The first three principal components for a data cube of Br$\beta$ (2.62 $\mu$m) of NGC 3256 N. The left panels show the tomograms of each component which describes the spatial intensity of each eigenspectrum. The black crosses show the peak continuum position at this wavelength. The right panels show the eigenspectrum of each principal component. The horizontal dashed lines show where the weight is zero and the vertical dashed lines show the kinematic centre, determined by the peak wavelength of the first principal component.
    The first component describes the rest frame emission, while the second captures rotation of the gas and the third caputures variations in the velocity dispersion.}
    \label{fig:PCADemo} 
\end{figure*}

We demonstrate this in Fig. \ref{fig:PCADemo}, with Br$\beta$ (2.62 $\mu$m) emission in NGC 3256 N, where the first principal component describes 83.5$\%$ of the variance of the data. Physically this corresponds to restframe emission of the line as the eigenspectrum shows a narrow line profile, centred on its restframe wavelength.  

The second component contains 13.3$\%$ of the variance in the data and appears to contain information about the rotation of the gas. Here the eigenspectrum shows excess flux redwards of the line centre and a lack of flux bluewards and thus describes a redshifted emission line. The tomogram shows the strength of this eigenspectrum across the galaxy where the greater positive weight suggests more redshifted emission while negative strengths describe the inverse i.e blueshifted emission. The second principal component here is consistent with the known rotation of this galaxy \citep[][]{Gomez2021}. Moreover, the wavelength where the eigenspectrum crosses from negative to positive is consistent with the restframe wavelength of the line and spatially coincides with the position of the nucleus. It is worth noting the tomogram is not a velocity map (see section \ref{sec:VelocityMaps}).

The third component describes 1.8$\%$ of the variance, and shows an eigenspectrum with excess flux in the wings where the tomogram is strongly positive and an excess flux with a narrow profile where the tomogram is strongly negative. This contributes more extreme values of velocity to the total velocity field. Subsequent principal components will have a diminishing contribution to the velocity field as the Scree test in Fig. \ref{fig:Scree} demonstrates. We show the effect of adding different numbers of principal components in the Supplementary Material, where components above three contribute minimally.

\subsection{Recovering Velocity Maps}
\label{sec:VelocityMaps}
One can reconstruct the original data cube by summing the eigensepctra where each is weighted by its corresponding tomogram. As the majority of the principal components simply describe noise in the data, reconstructing only using the first few components produces a data cube with less noise than the original data. A velocity map can then be constructed.

We reconstruct a data cube using only the first five components to obtain a velocity map. With this number of components, we are sure to capture all of the kinematics without the noise contained in the higher order components. 

As the first component describes the restframe emission, we take its eigenspectrum as a template for the rest profile of the emission feature. For each spaxel in the reconstructed cube, we calculate the wavelength (and thus velocity) shift required to match the restframe template to the emission for that spaxel. This provides a velocity for each spaxel and thus produces a velocity map. 

We compare the inferred velocity map against a more traditional method of fitting a Gaussian to each spaxel. We do this for all the emission lines used in this work and investigate the effect of the number of principal components in the reconstructed cube. This can be found in more detail in the Supplementary Material. We find that the PCA recovered map is typically consistent with the traditional method with at least the first three components required. For the PAH emission, we find that typically the first two components are sufficient.


\section{PAH Kinematics}
\label{sec:PAHKinematics}
\subsection{PCA of PAH Emission}
As outlined above, the PCA technique can isolate kinematic signatures by using the spatial distribution of the emission, without requiring any a priori knowledge of the emission profile. This technique is therefore well suited to detect/measure any kinematic signatures of the PAHs.

We focus on the 3.3 $\mu$m PAH as this feature is not heavily blended with other PAHs, is fairly symmetrical and has a high spatial resolution owing to NIRSpec observations. This feature is therefore ideal for this initial study of the kinematics of PAH features. Other PAH features, such as the 6.2 $\mu$m or 11.3 $\mu$m features, will be explored in future works.

\begin{figure*}
	\includegraphics[width=15cm]{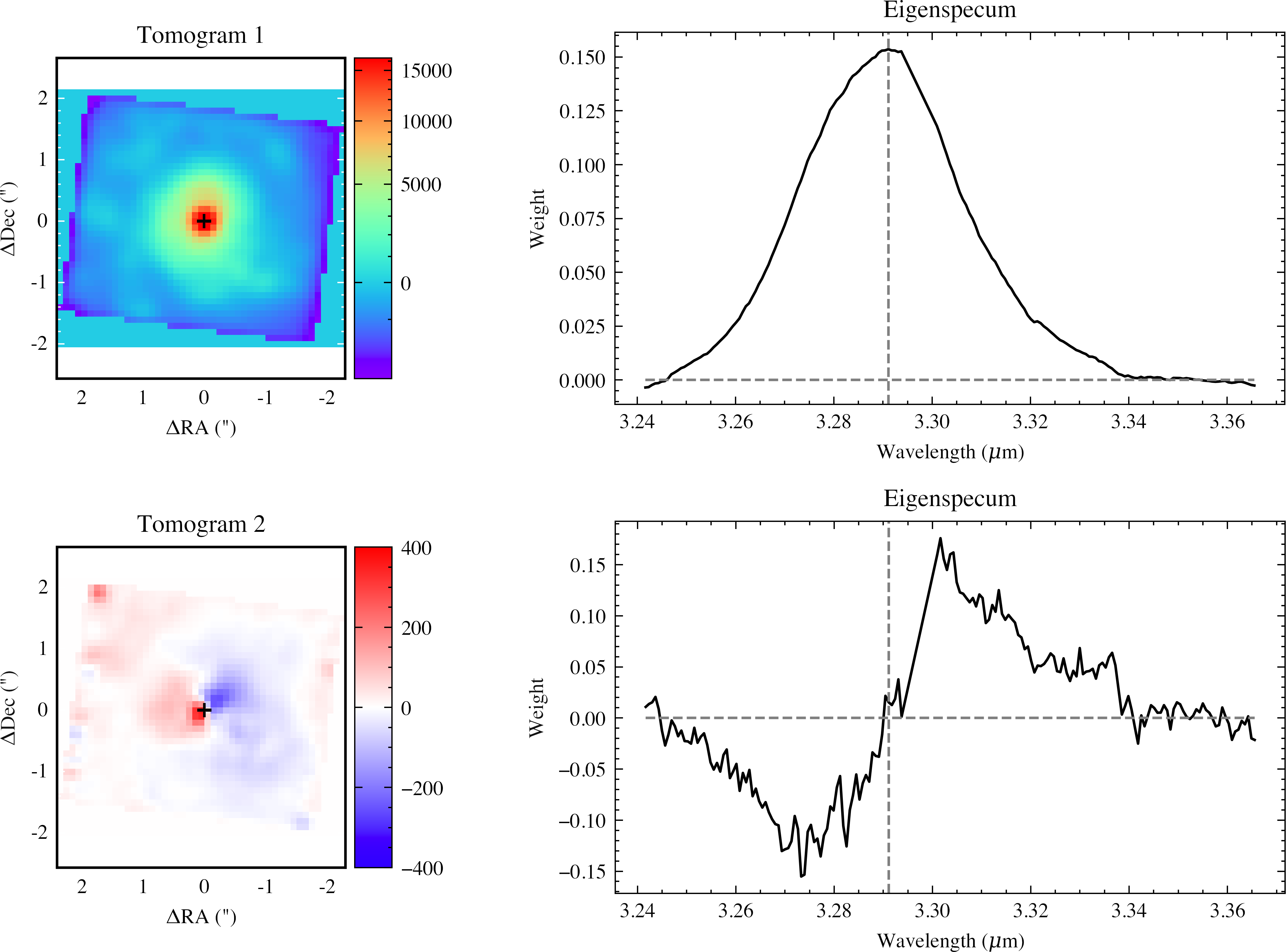}
    \caption{First two principal components of the 3.3 $\mu$m PAH in NGC 3256 N. The first component describes the restframe emission while the second describes kinematics of the PAHs. The left panels show the tomograms of each principal component, with the position of the continuum peak shown with a black cross. The right panels show the corresponding eigenspectra. }
    \label{fig:PAHTomograms} 
\end{figure*}

Before the PCA analysis, we mask the Pf$\delta$ (3.296 $\mu$m) line and subtract a local continuum to isolate the PAH feature. Fig. \ref{fig:PAHTomograms} shows the first two tomograms where the first describes the restframe emission and contains 99.9 $\%$ of the variance. The second principal component shows evidence of kinematics with the eigenspectrum showing excess flux redwards and a lack of flux bluewards, with the corresponding tomogram showing a gradient from positive to negative weights from east to west. The second component contains only 0.04$\%$ of the variance which highlights how burried the kinematic signature is within the data. Unlike Br$\beta$, the 3\textsuperscript{rd} principal component does not contain any useful information as more subtle changes in the profile, which would be captured by higher order components, are likely too difficult to detect in the data, unlike emission lines.

\begin{figure}
	\includegraphics[width=\columnwidth]{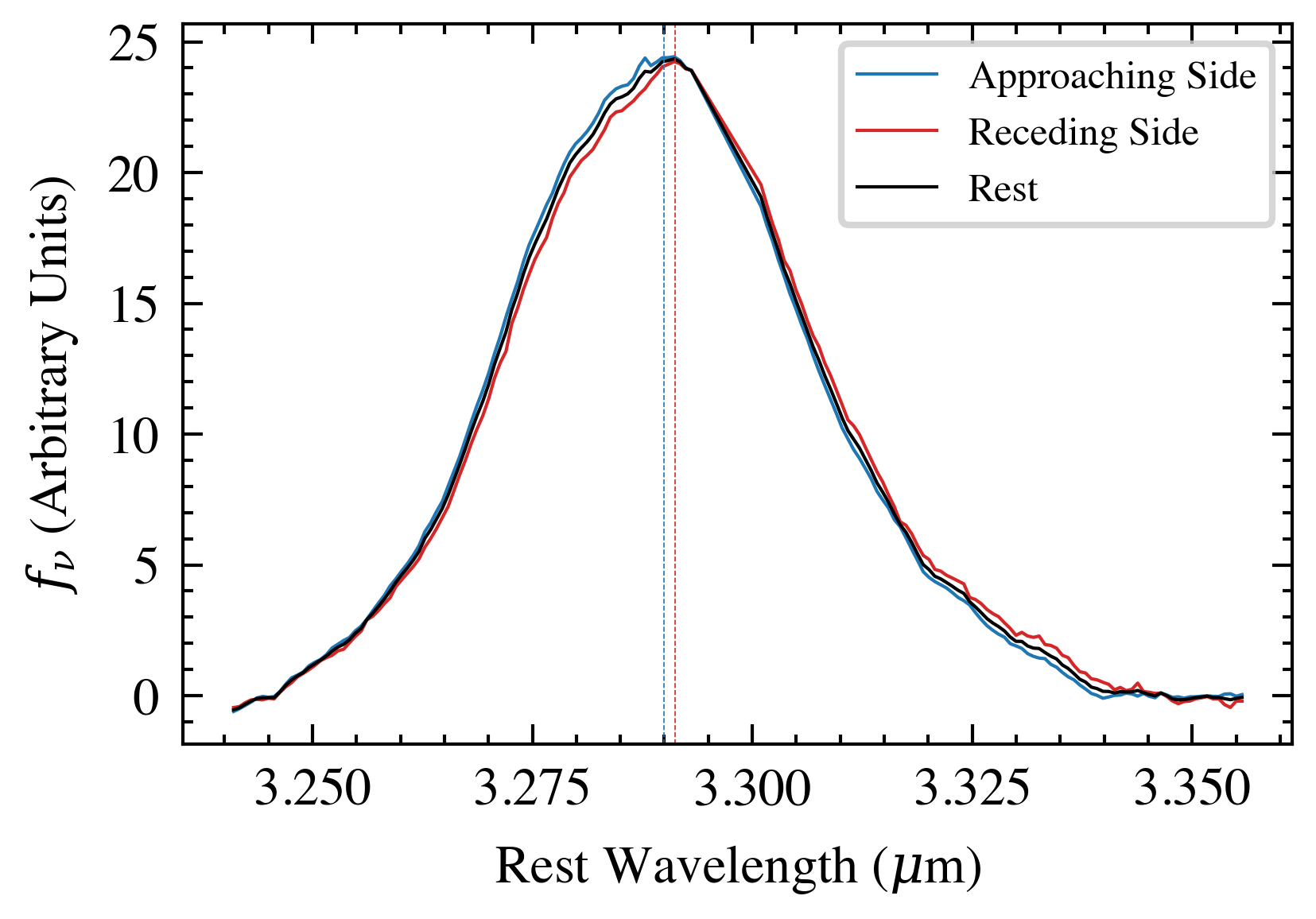}
    \caption{Average 3.3 $\mu$m PAH profiles for the approaching, receding and rest frame emission for NGC 3256 S. The vertical dashed lines show the mean wavelength of approaching and receding profiles. This plot highlights how small the velocity shifts are compared to the width of the emission feature, prompting the need to employ techniques such as PCA tomography.}
    \label{fig:PAHComp} 
\end{figure}

To demonstrate how small the velocity shifts are compared to the width of the PAH features, we plot the average PAH profile for the approaching and receding sides of the galaxy in Fig. \ref{fig:PAHComp}. This is calculated by taking the average of the emission feature in pixels where the second tomogram is only positive (receding) or only negative (approaching).

For each of the three targets we apply the PCA method to Br$\beta$ (2.62 $\mu$m) to trace the kinematics of young stars, H$_2$ 1-0 S(1) (2.12 $\mu$m) to trace molecular gas, and the 3.3 $\mu$m PAH to trace dust, as discussed in the previous section. We then compare the kinematics of these tracers. 

\begin{figure*}
	\includegraphics[width=14cm]{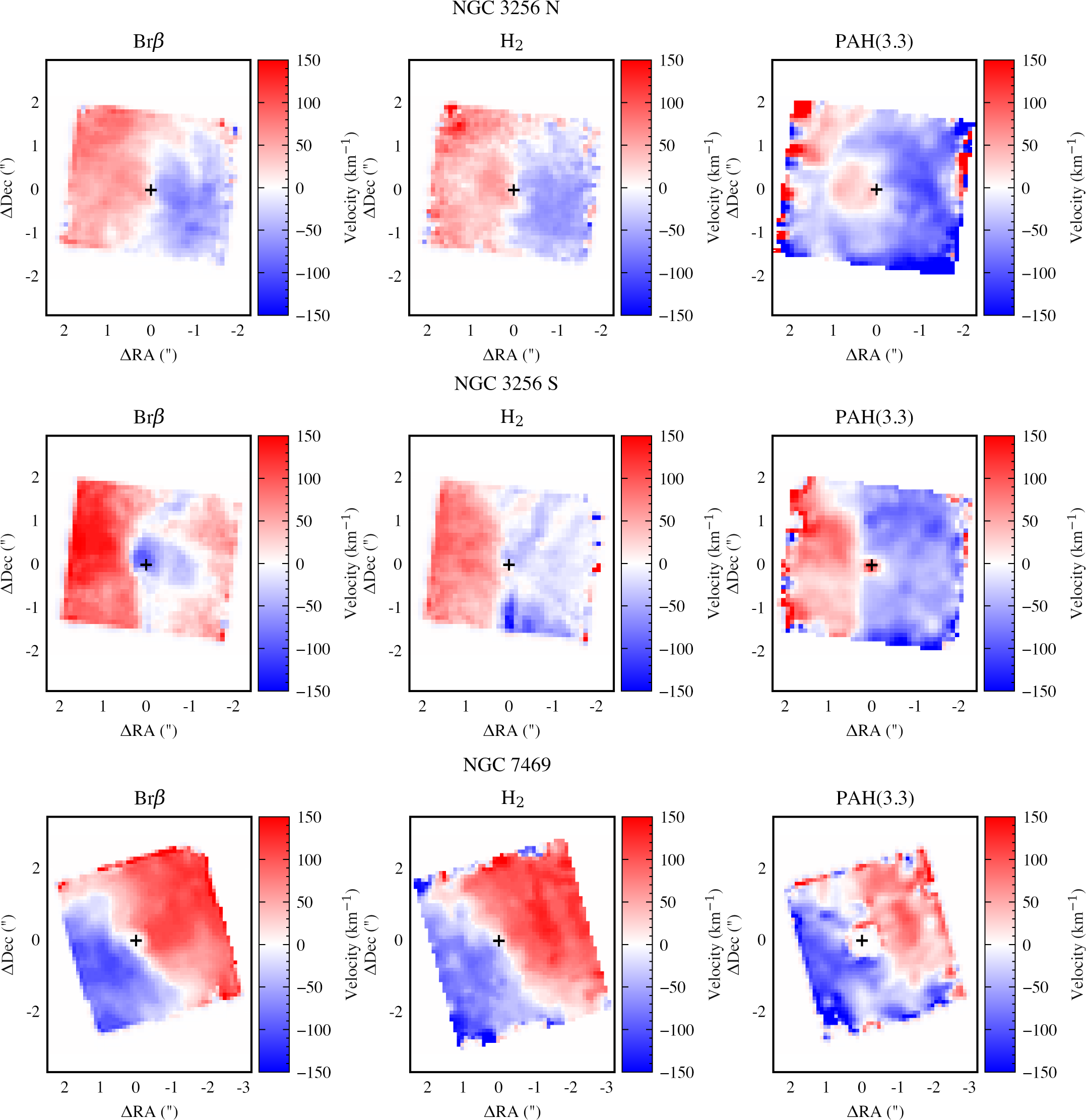}
    \caption{Velocity maps inferred from the PCA decomposition for the three targets. In each case the velocity map is constructed using the first five principal components. The left most panels show Br$\beta$ (2.625 $\mu$m), the middle panels show H$_2$ 1-0 S(1) (2.12 $\mu$m) and the right panels show the 3.3 $\mu$m PAH maps. For all the maps, the black cross denotes the continuum peak position at the wavelengths of the emission line.}
    \label{fig:VelMaps} 
\end{figure*}

Fig. \ref{fig:VelMaps} shows the velocity maps for each target with each emission feature. Before applying the PCA method to the PAH cubes, we smooth each channel with a Gaussian filter with a $\sigma = 1$ pixel. This is due to the low signal to noise of the kinematics in the data. We additionally do this for the Br$\beta$ and H$_2$ lines for NGC 7469, as the signal to noise is lower for this target. We also mask the point source in NGC 7469 as the PSF is particularly strong for this target, which introduces ``wiggles'' into individual pixels near the centre \citep[e.g.][]{Perna2023}, disrupting the PCA decomposition.

\section{Discussion}

For all three targets, the PAHs show clear signatures of kinematics and qualitatively follow the rotation of the galaxy. We calculate a Pearson correlation coefficient between the velocity maps of the various tracers to quantify how well the different emission features trace the same kinematics. This is shown in Table. \ref{tab:Correlations}.

NGC 7469 shows the strongest agreement between the gas tracers and the PAHs both qualitatively as well as having the highest Pearson correlation coefficients. This is despite the presence of a nuclear outflow \citep[e.g.][]{Garcia-Bernete2022c, Armus2023}, although the outflow is most prominent in the ionised phase. For a more detailed kinematic analysis for this target see \citet{Bianchin2023, Vivian2022}. 

While the other two targets also broadly show agreement, there are some clear differences between the velocity maps. For NGC 3256 S, the H$_2$ maps correlates well with the PAHs however Br$\beta$ shows a peculiar velocity field and thus the correlation is worse. This is in agreement with the maps shown in \citet{Pereira-Santaella2023, Pereira-Santaella2024}, however the reason for the unusual velocity field of Br$\beta$ is unclear.
It is also worth noting that the velocity map of H$_2$ is not dominated by the strong/collimated molecular outflow driven by the obscured AGN \citep[e.g.][]{Sakamoto2014, Pereira-Santaella2023}. This is likely due to the high inclination of the galaxy where the line of sight velocity is dominated by rotation whereas the radial component of the outflow will be relatively weak. 

For NGC 3256 N, the kinematic position angle of the PAH velocity map appears to be different to Br$\beta$ and H$_2$, resulting in a poorer correlation, which may suggest a warp in the inner disk. This can be clearly seen in the eigenspectrum of the 2\textsuperscript{nd} component in Fig. \ref{fig:PAHTomograms}, which is dominated by the rotation. In addition there appears to be more blueshifted emission which may be due to the rest emission being estimated incorrectly. As we used the 1\textsuperscript{st} principal component as the rest emission, this provides an average over the feild of view. If for example, more of one side of the disk was visible in the field of view, this would shift the rest emission as given by the 1\textsuperscript{st} principal component. In this work however, this is not a likely explanation as the nuclei are centered in the observations.

\begin{table}
\centering
  \caption{Velocity Map Correlations. The 4\textsuperscript{th} column shows the Pearson correlation coefficient between the two given maps.}
  \label{tab:Correlations}
    \def\arraystretch{1.}
    \setlength{\tabcolsep}{10pt}
    \begin{threeparttable}
  \begin{tabular}{cccc}
  
    \hline

     Galaxy & Correlation & $N$\textsubscript{spaxels} & $\rho$ \\
        \hline

NGC 3256 N & Br$\beta$ vs H$_2$ & 1490 & 0.89 \\
NGC 3256 N & PAH(3.3) vs H$_2$ & 1490 & 0.24 \\
NGC 3256 N & PAH(3.3) vs Br$\beta$ & 1650 & 0.25 \\
NGC 3256 S & Br$\beta$ vs H$_2$ & 1478 & 0.75 \\
NGC 3256 S & PAH(3.3) vs H$_2$ & 1494 & 0.68 \\
NGC 3256 S & PAH(3.3) vs Br$\beta$ & 1597 & 0.60 \\
NGC 7469 & Br$\beta$ vs H$_2$ & 2193 & 0.86 \\
NGC 7469 & PAH(3.3) vs H$_2$ & 2193 & 0.74 \\
NGC 7469 & PAH(3.3) vs Br$\beta$ & 2426 & 0.80 \\

    \hline

  \end{tabular}
  \end{threeparttable}
 \end{table}


\section{Conclusions}
In this letter we have demonstrated how the kinematics of PAHs can be inferred using PCA tomography. Our main findings are

\begin{itemize}
    \item We demonstrate the technique of PCA tomography with JWST/NIRSpec data, where the second component isolates the rotation of the gas while the third corresponds to the velocity dispersion. 
    \item By applying the PCA technique, we detect a kinematic signature of the 3.3 $\mu$m PAH feature in all three targets. 
    \item By reconstructing a velocity map, we find the PAH emission to follow the rotation of the galaxy, with consistent values of velocity. 
\end{itemize}
With this technique, future work exploring the kinematics of other PAH features, by incorporating MIRI MRS data, will provide a new window into this crucial component of the ISM.

\section*{Acknowledgements}
The authors are grateful to the DD-ERS team (Program 1328, PI: Lee Armus \& Aaron Evans) for the observing program with a zero–exclusive–access period.
FRD acknowledges support from STFC through grant ST/W507726/1. DR and IGB acknowledge support from STFC through grant ST/S000488/1 and ST/W000903/1.

\section*{Data Availability}
All the data used in this work is publicly available as part of DD-ERS Program 1328, downloadable from the \href{https://mast.stsci.edu/portal/Mashup/Clients/Mast/Portal.html}{MAST archive}.



\bibliographystyle{mnras}
\bibliography{References} 

\bsp	
\label{lastpage}
\end{document}


\label{firstpage}
\pagerange{\pageref{firstpage}--\pageref{lastpage}}
\maketitle





\section{Number of PCA Components}
We described in the main text how we generate velocity maps from the components obtained from the PCA decomposition. In this section we demonstrate how the inferred velocity map changes based on the number of components included in the reconstructed cube. We then compare this to a traditional method of modelling the line profile of each spaxel in the original data cube. We do this for each of the emission lines.

For the traditional method we fit a Gaussian to the continuum subtracted spectra from each pixel. The velocity is inferred by measuring the difference in wavelength of the centre of the fitted Gaussian to the restframe centre wavelength of the emission line. 

\begin{figure*}
	\includegraphics[width=\textwidth]{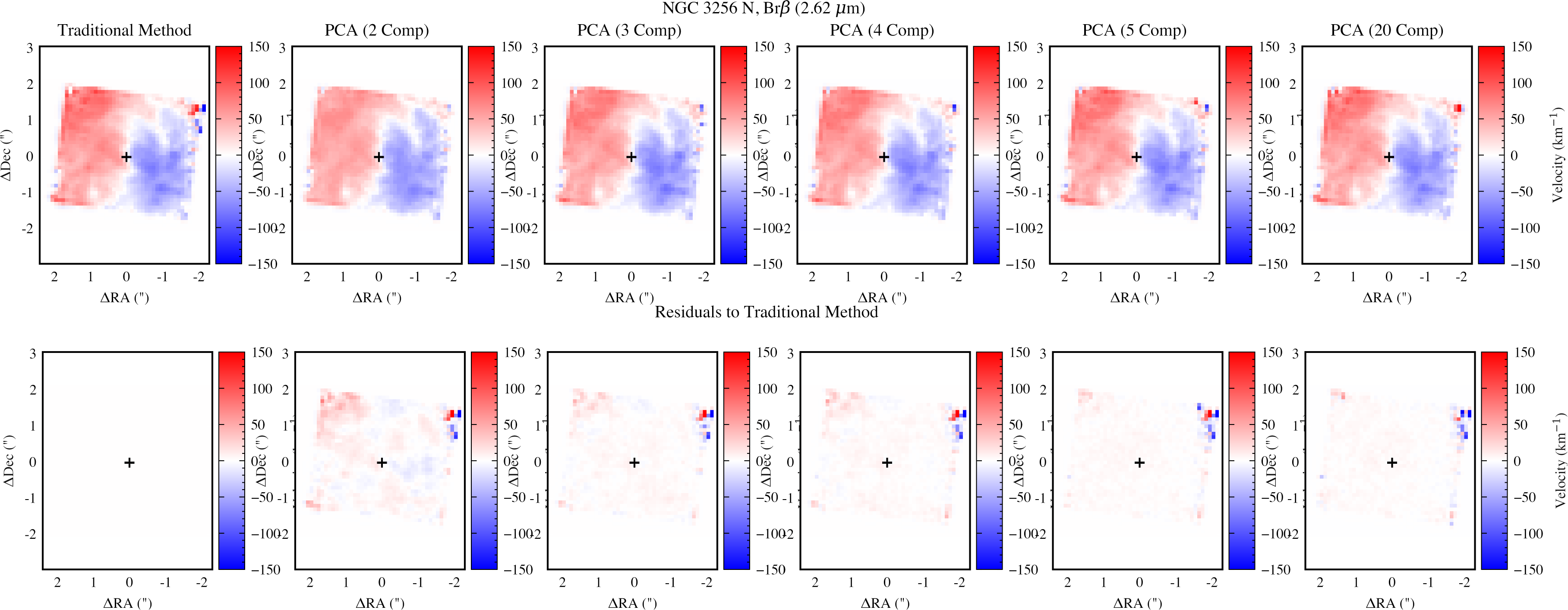}
    \caption{Comparison between the inferred velocity maps from the traditional method of fitting a Gaussian to each spaxel and those from the PCA, with different number of components. The top panels show the velocity maps while the lower panels show the residuals to the traditional method. }
    \label{fig:NGC3256NBrB_Comp} 
\end{figure*}

We first demonstrate this for Br$\beta$ in NGC 3256 N. We find that the first two principal components contain the vast majority of the information about the velocity field, with subsequent principal components contain ever decreasing information. The higher order components appear to contain information about the more extreme values of velocity in the field which presents as smaller variations in the line profiles. We show in Fig. \ref{fig:NGC3256NBrB_Comp}, the comparison between the traditional method and the PCA reconstructed velocity map from different number of components. The PCA velocity maps from just the first two components appears too ``flat'', while including the third component shows a closer match to the traditional method. Beyond the third component, very little changes as most of the information is contained in the first three components, as learnt from the Scree test.

\begin{figure*}
	\includegraphics[width=\textwidth]{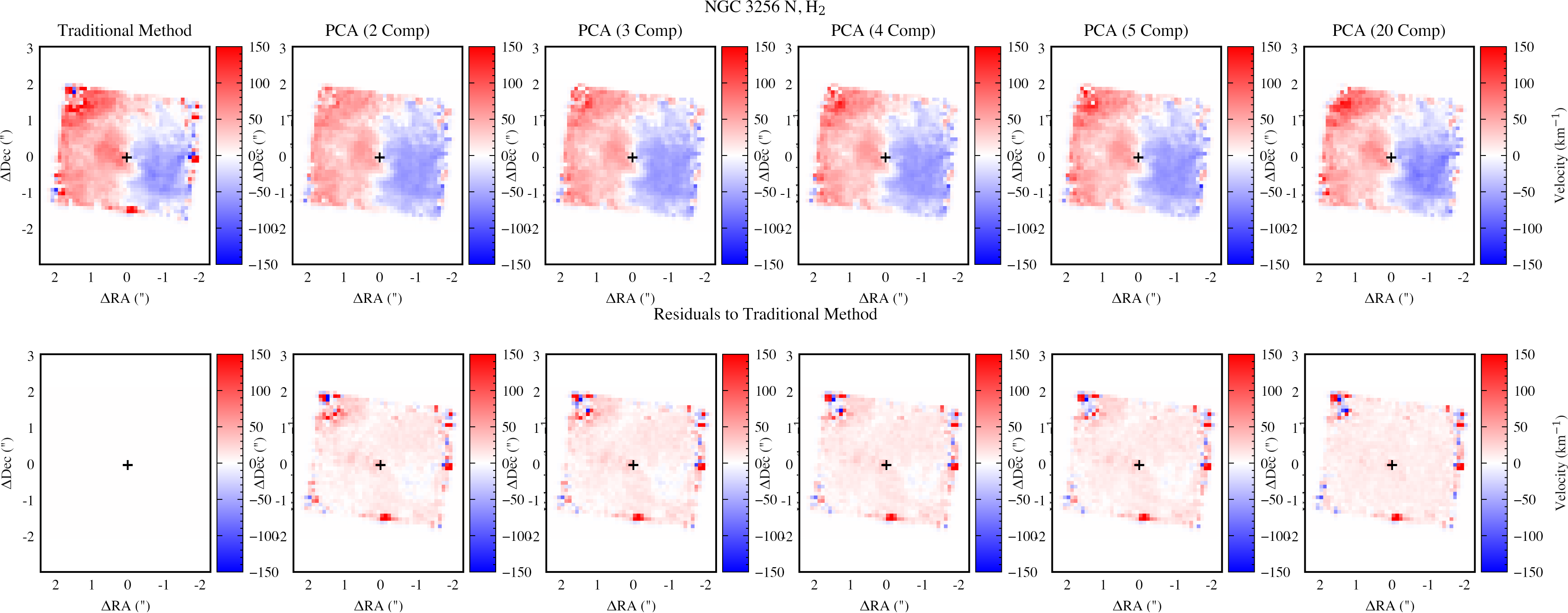}
    \caption{Same as Fig. \ref{fig:NGC3256NBrB_Comp} but for NGC 3256 N H$_2$.}
    \label{fig:NGC3256NH2_Comp} 
\end{figure*}

\begin{figure*}
	\includegraphics[width=\textwidth]{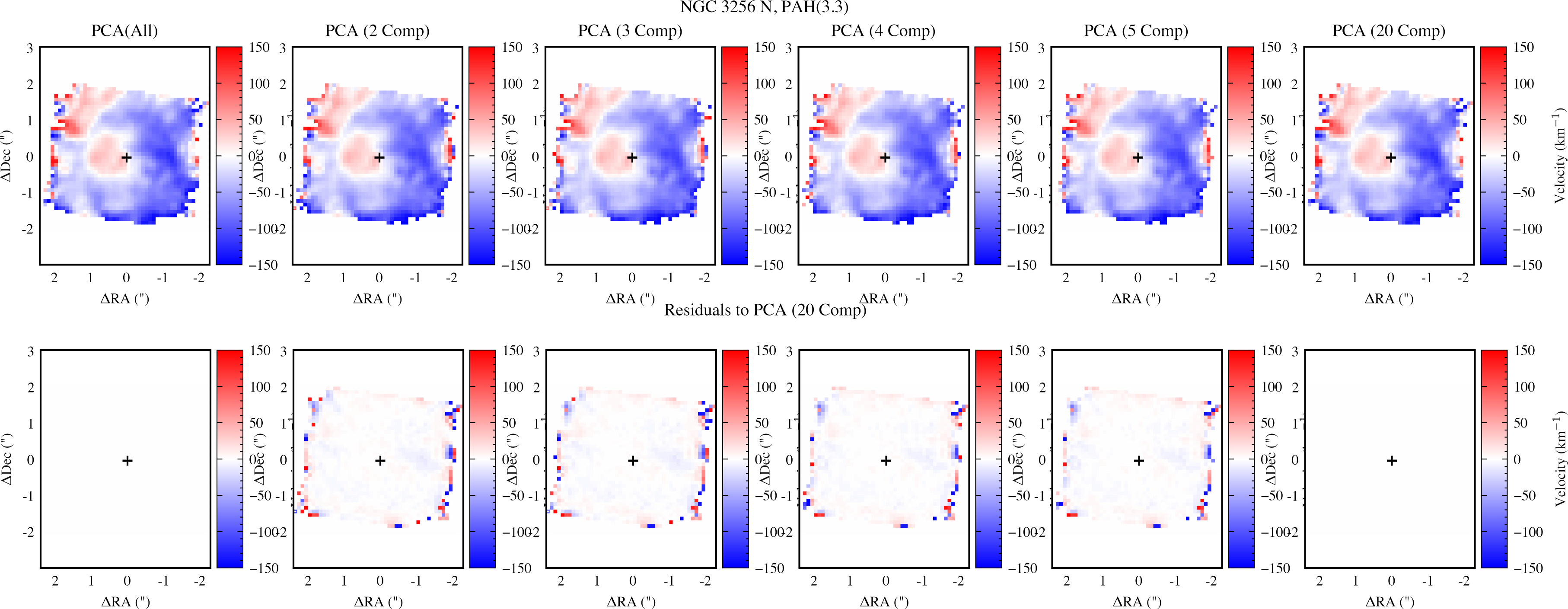}
    \caption{Same as Fig. \ref{fig:NGC3256NBrB_Comp} but for the 3.3 $\mu$m PAH emission in NGC 3256 N. As the traditional method is not suitable for PAHs, we take the residuals relative to the 20 component PCA map.}
    \label{fig:NGC3256NPAH_Comp} 
\end{figure*}

\begin{figure*}
	\includegraphics[width=\textwidth]{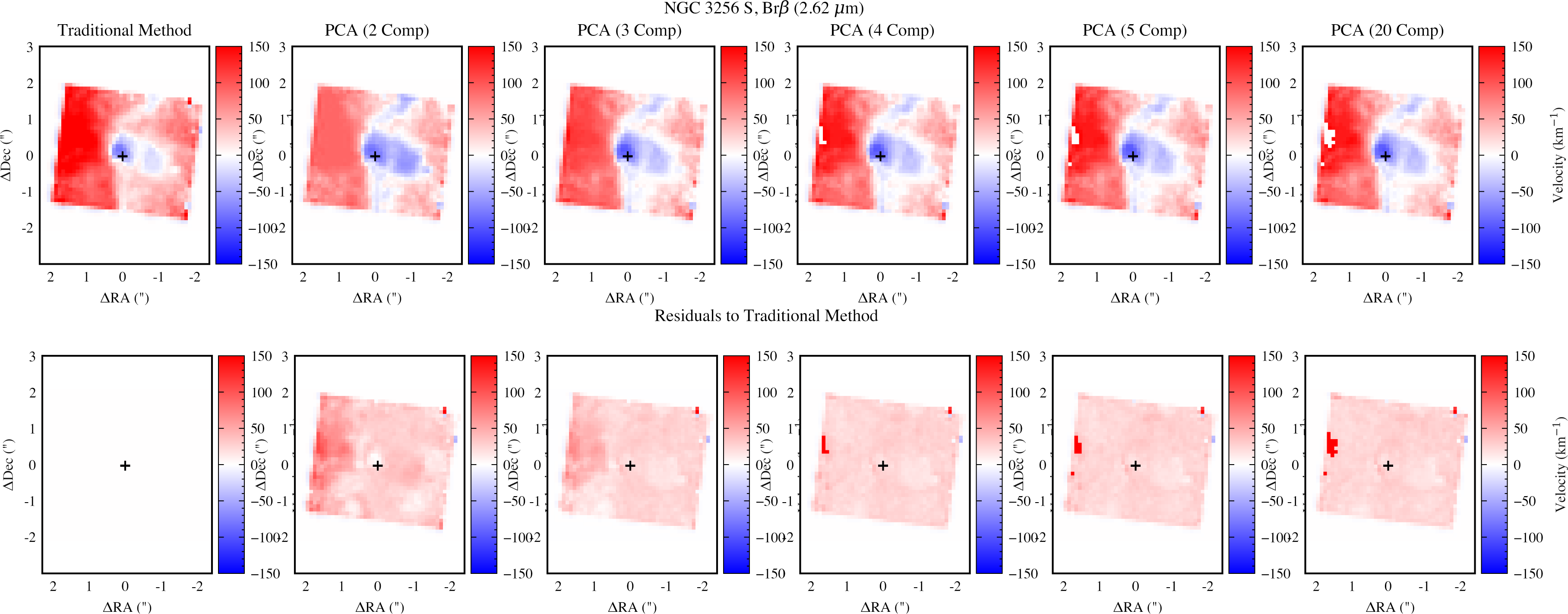}
    \caption{Same as Fig. \ref{fig:NGC3256NBrB_Comp} but for NGC 3256 S Br$\beta$.}
    \label{fig:NGC3256SBrb_Comp} 
\end{figure*}

\begin{figure*}
	\includegraphics[width=\textwidth]{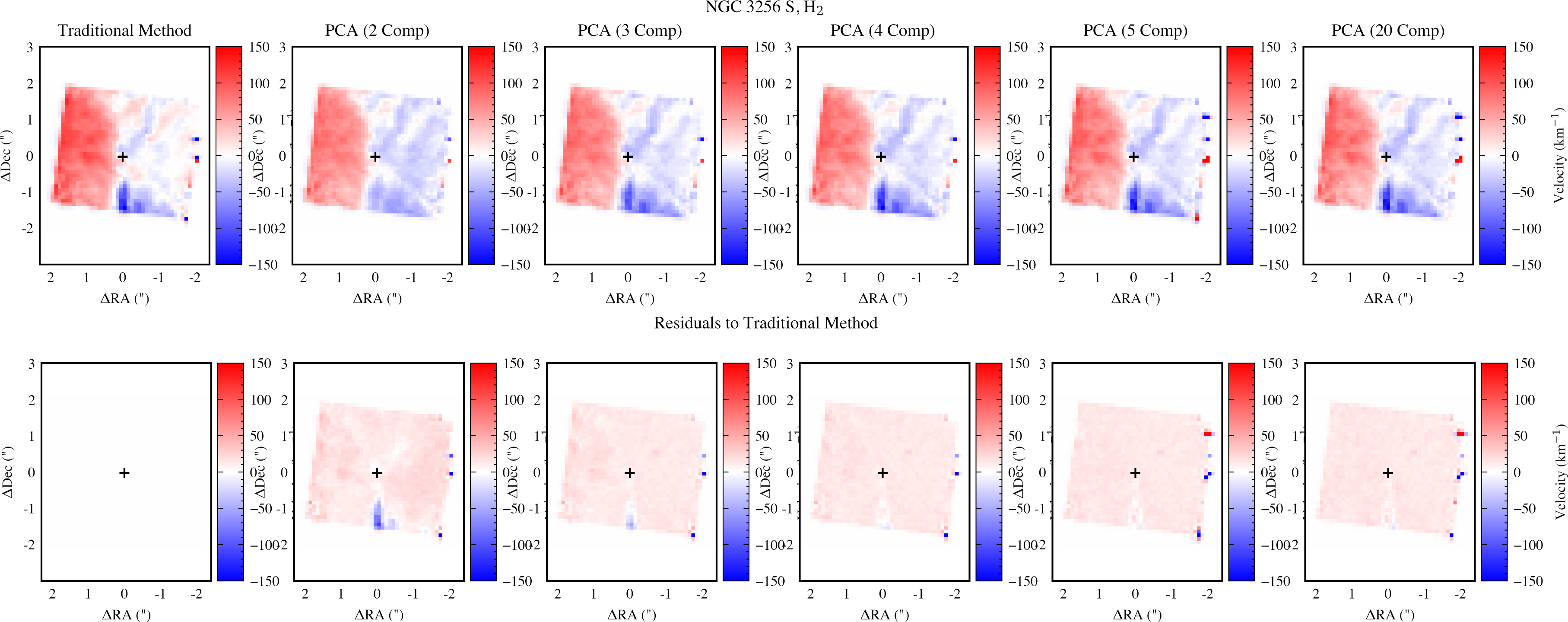}
    \caption{Same as Fig. \ref{fig:NGC3256NBrB_Comp} but for NGC 3256 S H$_2$.}
    \label{fig:NGC3256SH2_Comp} 
\end{figure*}

\begin{figure*}
	\includegraphics[width=\textwidth]{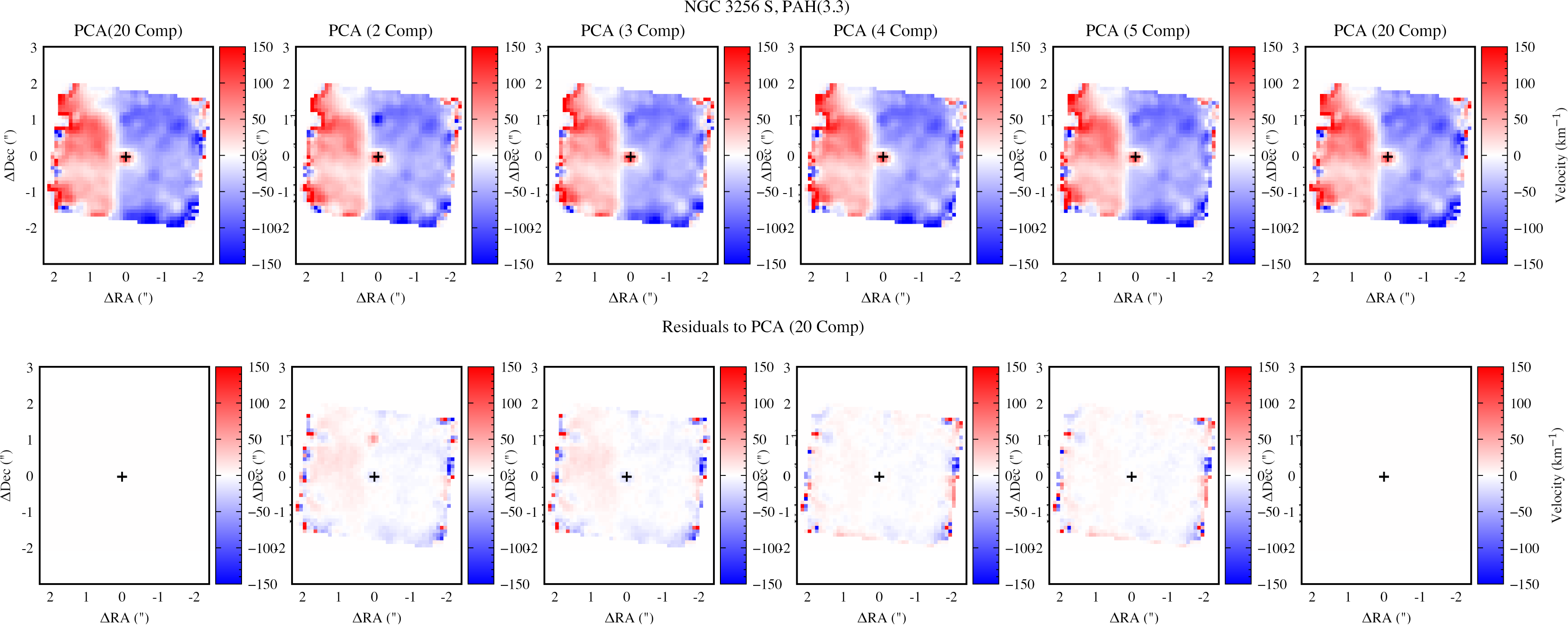}
    \caption{Same as Fig. \ref{fig:NGC3256NBrB_Comp} but for the 3.3 $\mu$m PAH emission in NGC 3256 S. As the traditional method is not suitable for PAHs, we take the residuals relative to the 20 component PCA map.}
    \label{fig:NGC3256SPAH_Comp} 
\end{figure*}

\begin{figure*}
	\includegraphics[width=\textwidth]{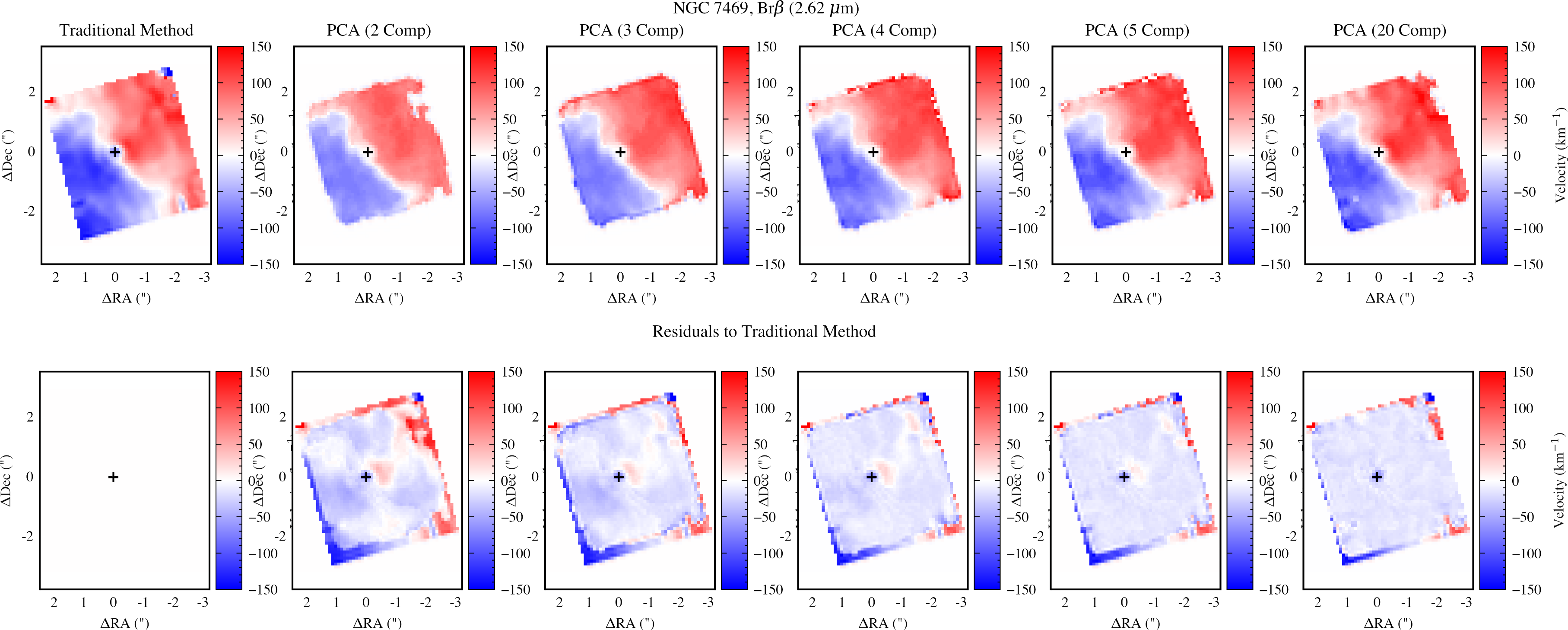}
    \caption{Same as Fig. \ref{fig:NGC3256NBrB_Comp} but for NGC 7469 Br$\beta$.}
    \label{fig:NGC7469Brb_Comp} 
\end{figure*}

\begin{figure*}
	\includegraphics[width=\textwidth]{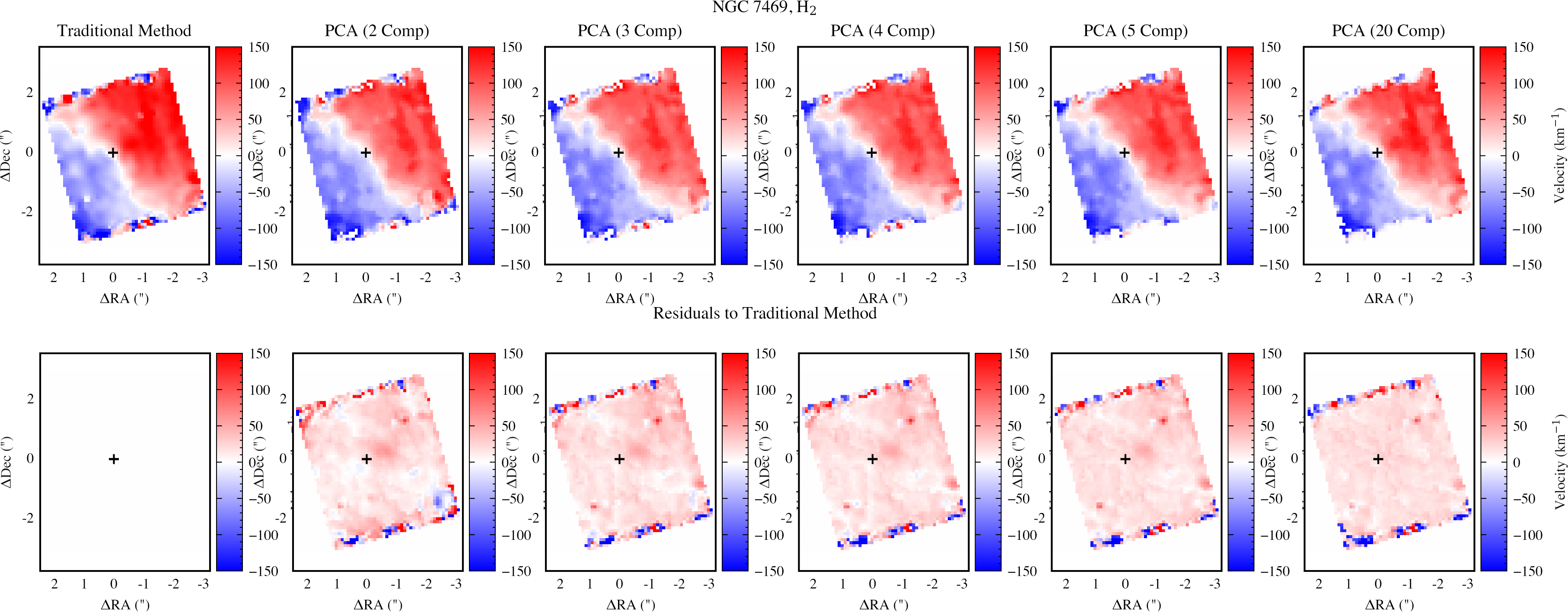}
    \caption{Same as Fig. \ref{fig:NGC3256NBrB_Comp} but for NGC 7469 H$_2$.}
    \label{fig:NGC7469H2_Comp} 
\end{figure*}

\begin{figure*}
	\includegraphics[width=\textwidth]{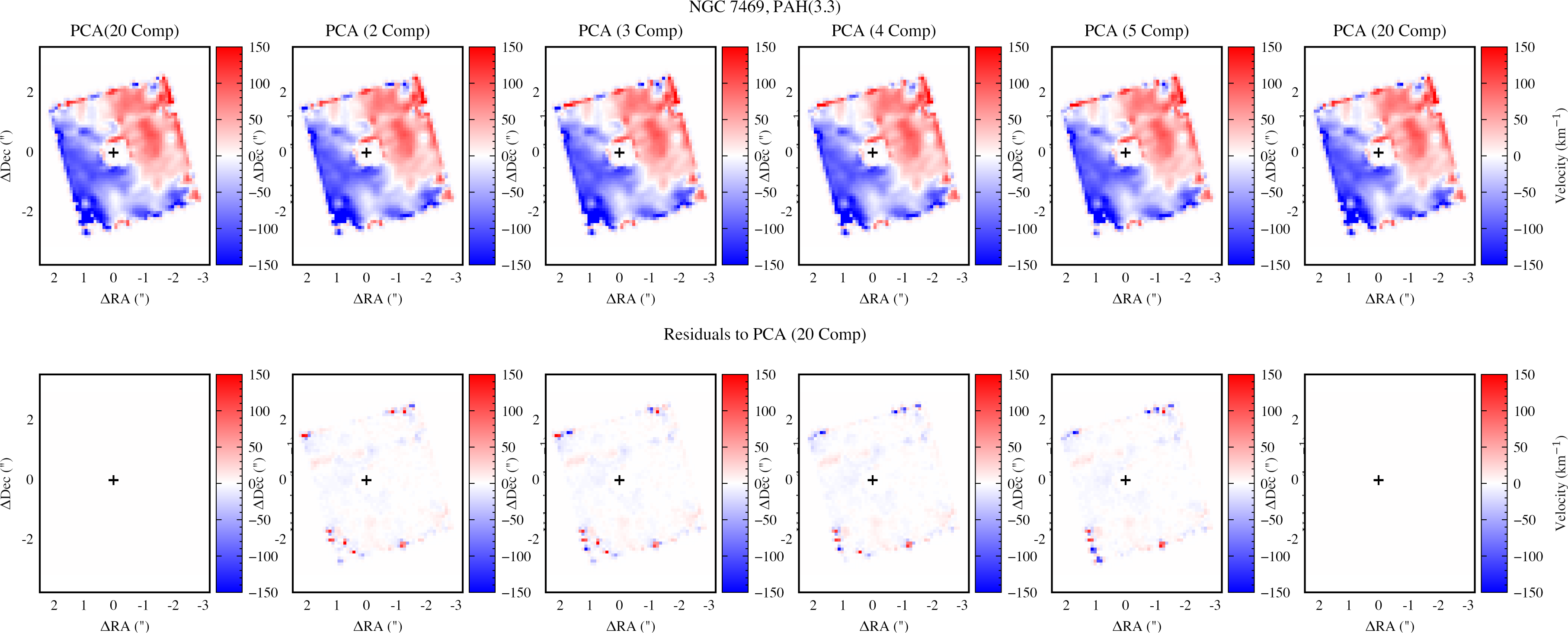}
    \caption{Same as Fig. \ref{fig:NGC3256NBrB_Comp} but for the 3.3 $\mu$m PAH emission in NGC 7469. As the traditional method is not suitable for PAHs, we take the residuals relative to the 20 component PCA map.}
    \label{fig:NGC7469PAH_Comp} 
\end{figure*}

We additionally show the same comparison for the remaining lines/PAHs used in this work in Figs. \ref{fig:NGC3256NH2_Comp}, \ref{fig:NGC3256NPAH_Comp}, \ref{fig:NGC3256SBrb_Comp}, \ref{fig:NGC3256SH2_Comp}, \ref{fig:NGC3256SPAH_Comp}, \ref{fig:NGC7469Brb_Comp}, \ref{fig:NGC7469H2_Comp}, \ref{fig:NGC7469PAH_Comp}.


\section{Additional PCA Plots}
\begin{figure*}
	\includegraphics[width=\textwidth]{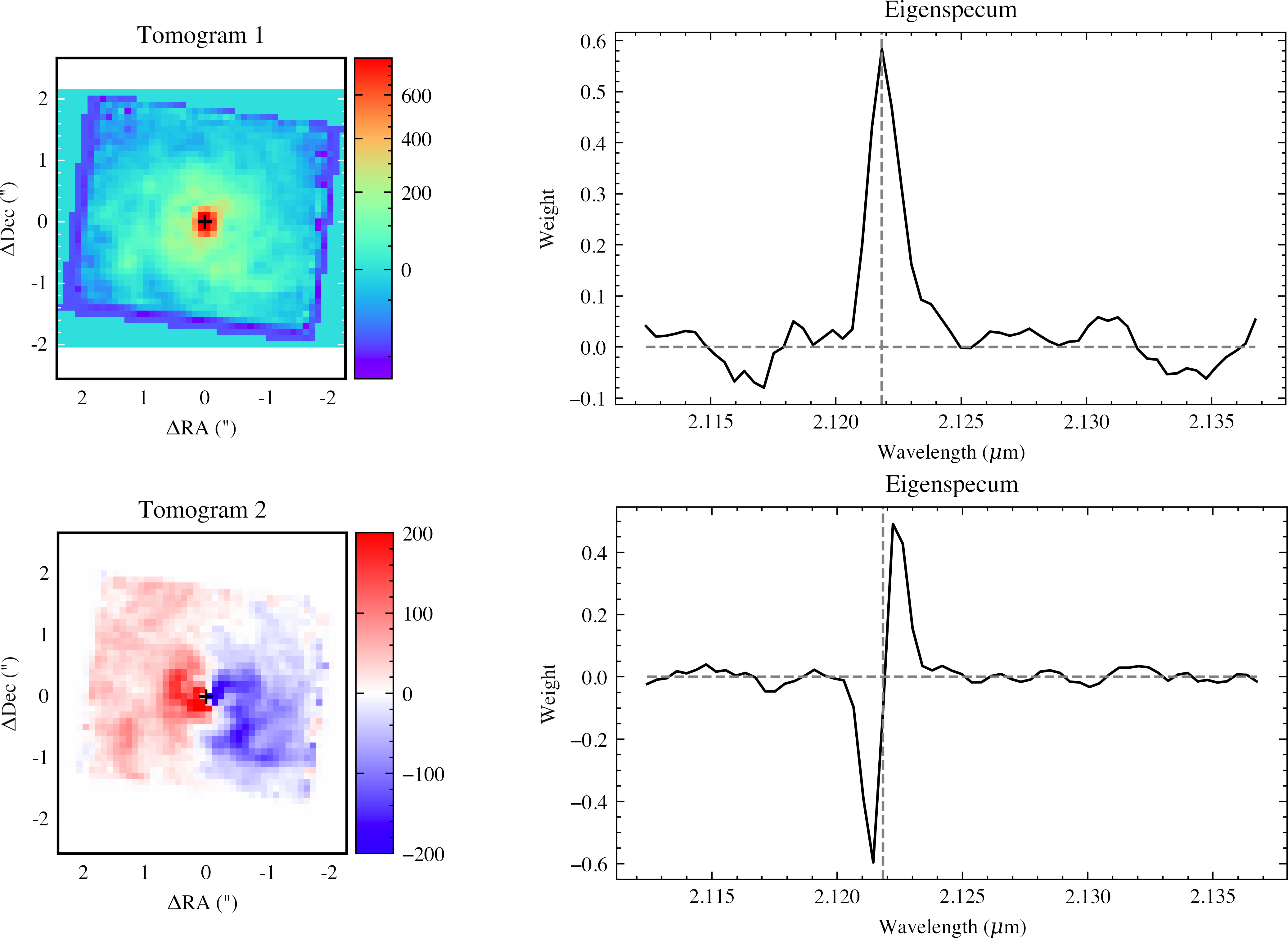}
    \caption{The first two principal components for a data cube of H$_2$ of NGC 3256 N. The left panels show the tomograms of each component which describes the spatial intensity of each eigenspectrum. The black crosses show the peak continuum position at this wavelength. The right panels show the eigenspectrum of each principal component. The horizontal dashed lines show where the weight is zero and the vertical dashed lines show the kinematic centre, determined by the peak wavelength of the first principal component.
    The first component describes the rest frame emission, while the second captures rotation of the gas.}
    \label{fig:NGC3256NH2} 
\end{figure*}

\begin{figure*}
	\includegraphics[width=\textwidth]{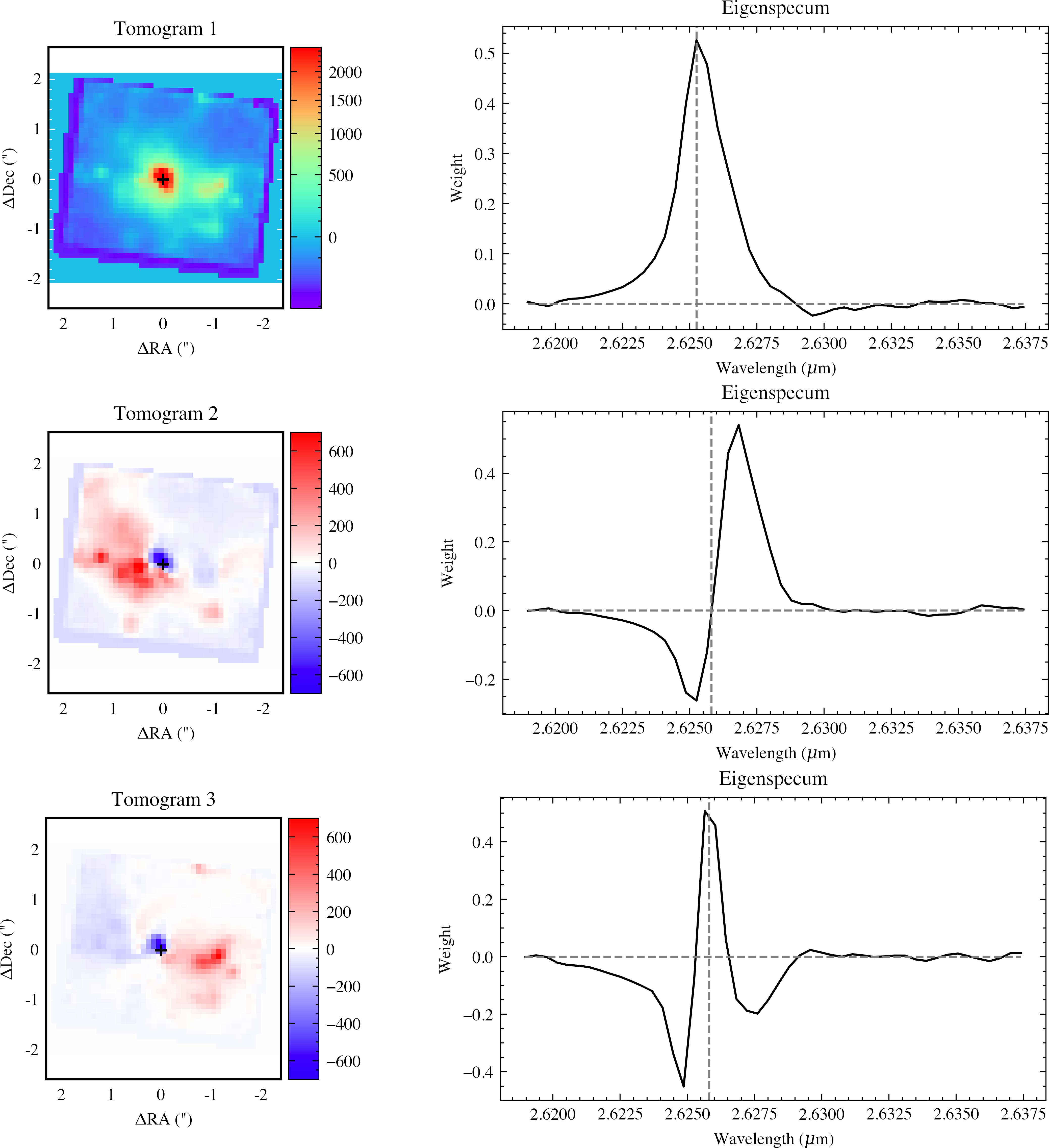}
    \caption{The first three principal components for a data cube of Br$\beta$ of NGC 3256 S. The left panels show the tomograms of each component which describes the spatial intensity of each eigenspectrum. The black crosses show the peak continuum position at this wavelength. The right panels show the eigenspectrum of each principal component. The horizontal dashed lines show where the weight is zero and the vertical dashed lines show the kinematic centre, determined by the peak wavelength of the first principal component.
    The first component describes the rest frame emission, while the second captures rotation of the gas and the third describes the velocity dispersion.}
    \label{fig:NGC3256SBrB} 
\end{figure*}

\begin{figure*}
	\includegraphics[width=\textwidth]{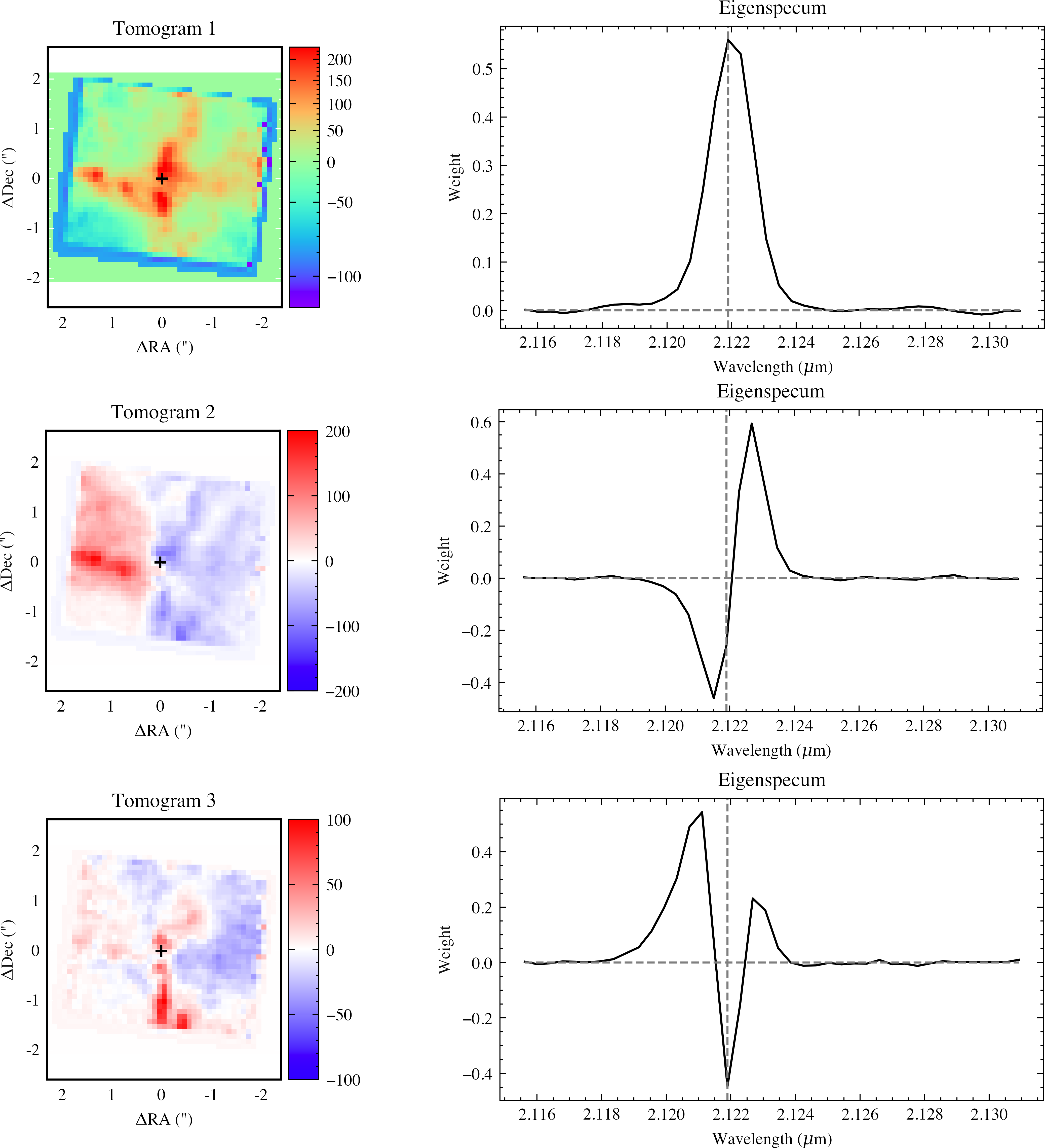}
    \caption{Same as Fig. \ref{fig:NGC3256SBrB} but for H$_2$ in NGC 3256 S.}
    \label{fig:NGC3256SH2} 
\end{figure*}

\begin{figure*}
	\includegraphics[width=\textwidth]{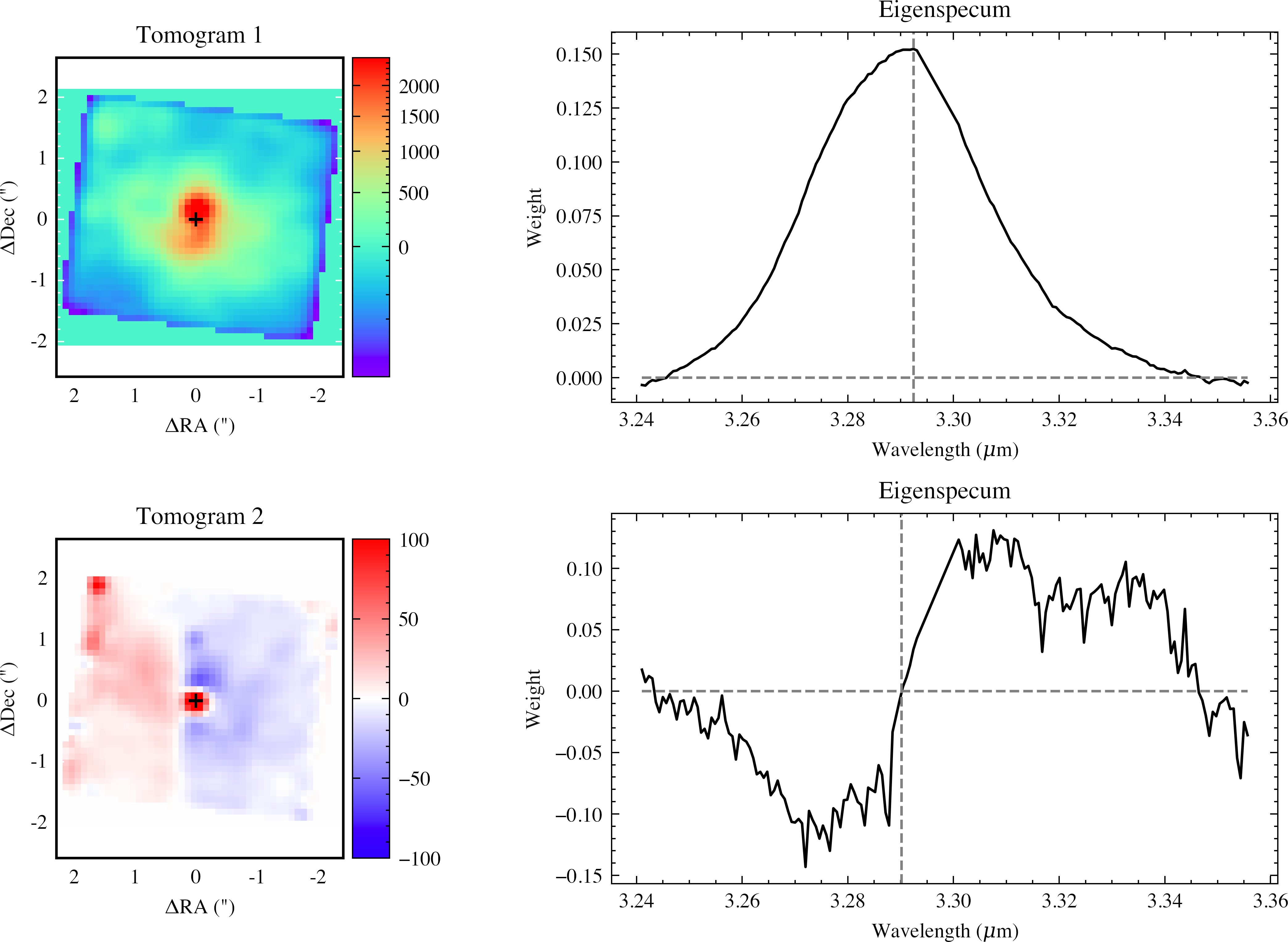}
    \caption{Same as Fig. \ref{fig:NGC3256SBrB} but for the 3.3 $\mu$m PAH in NGC 3256 S.}
    \label{fig:NGC3256SPAH} 
\end{figure*}

\begin{figure*}
	\includegraphics[width=15cm]{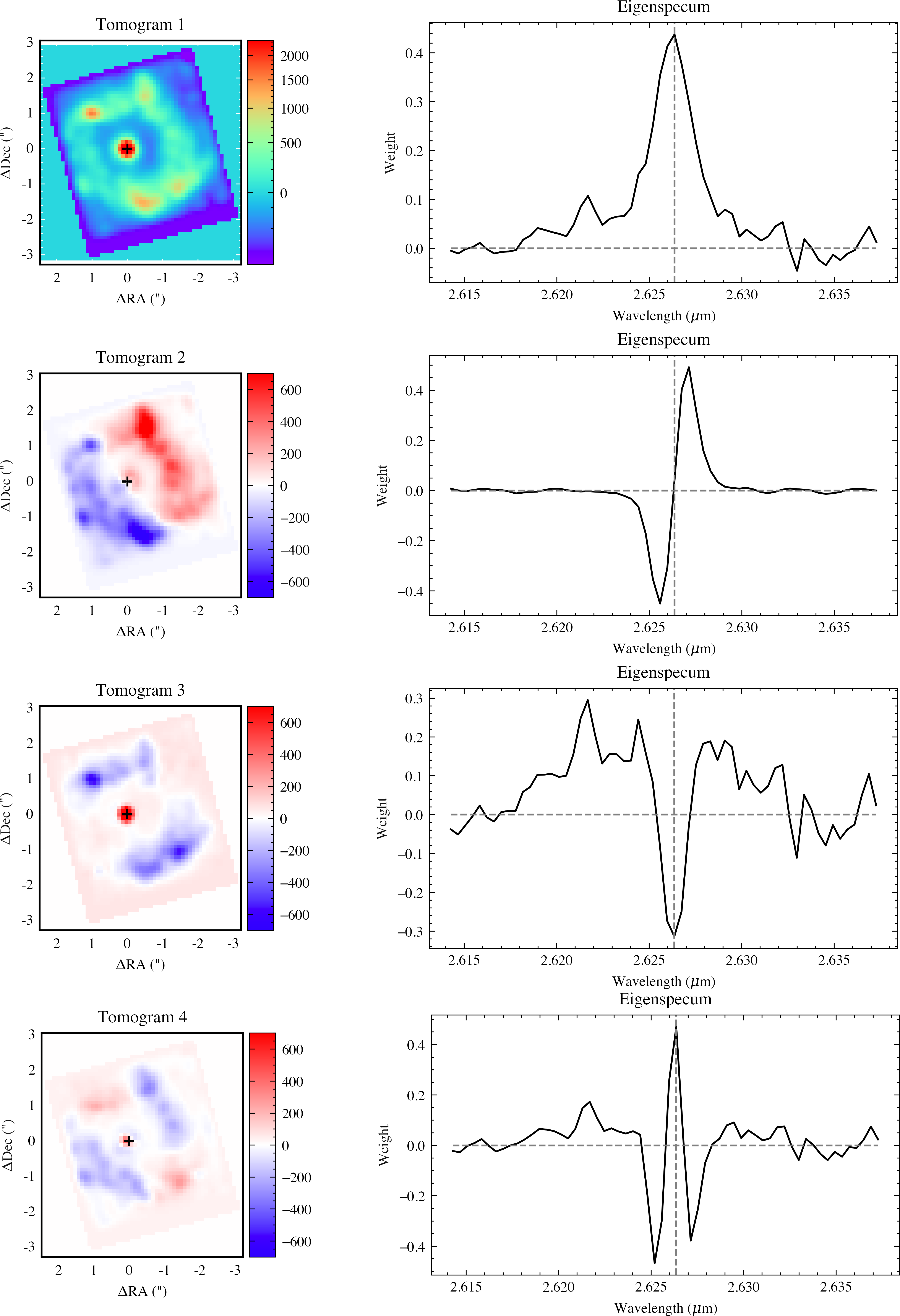}
    \caption{The first four principal components for a data cube of Br$\beta$ of NGC 7469. The first component describes the rest frame emission, while the second captures rotation of the gas. The third and fourth both show different components of the velocity dispersion, with the third isolating the broad line region of the AGN, with the very broad wings. The fourth component shows a higher velocity dispersion in the north-west and south-east of the star-forming ring, coinciding with the nuclear outflow.}
    \label{fig:NGC7469BrB} 
\end{figure*}

\begin{figure*}
	\includegraphics[width=\textwidth]{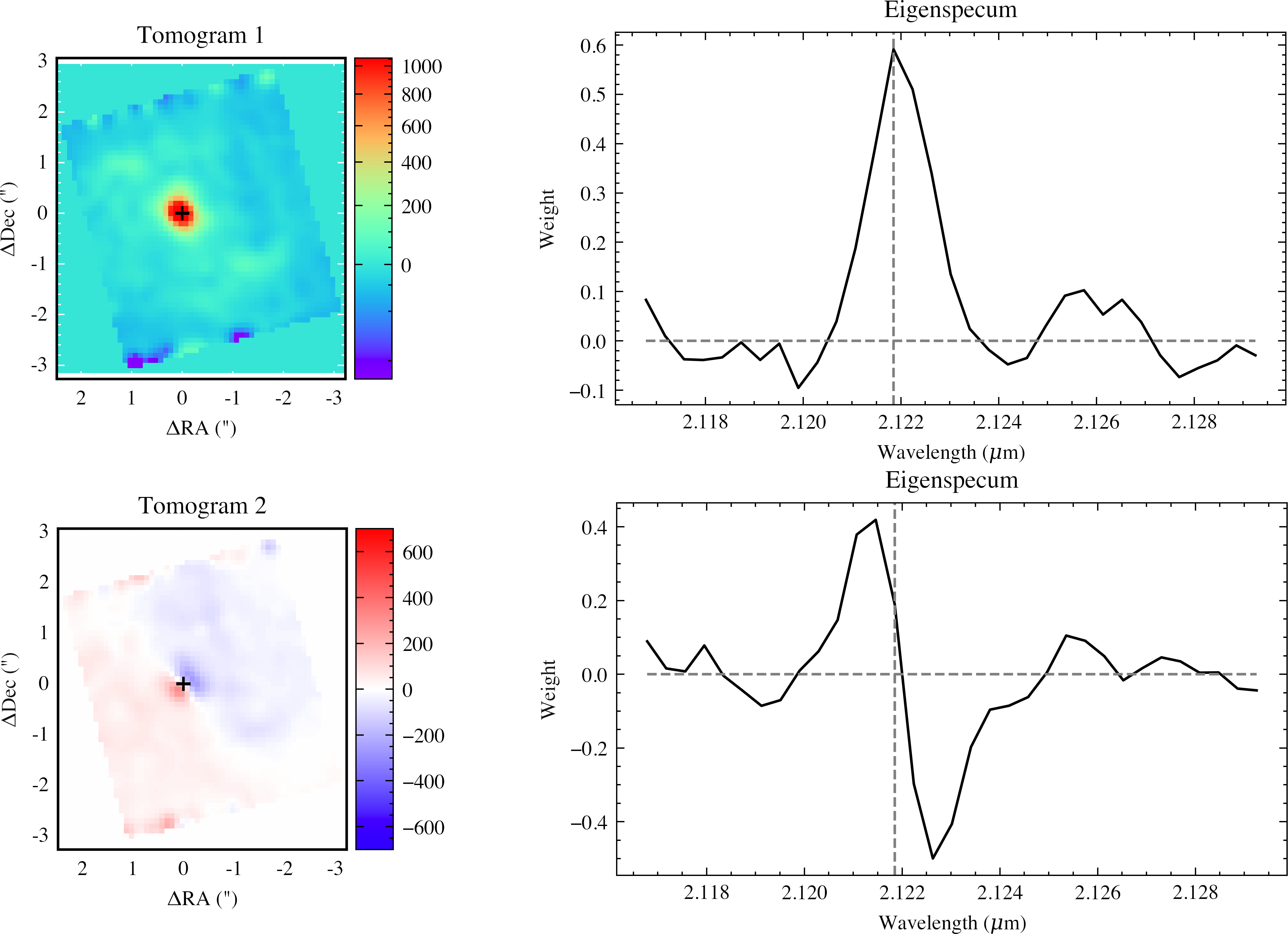}
    \caption{Same as Fig. \ref{fig:NGC3256SBrB} but for H$_2$ in NGC 7469.}
    \label{fig:NGC7469H2} 
\end{figure*}

\begin{figure*}
	\includegraphics[width=\textwidth]{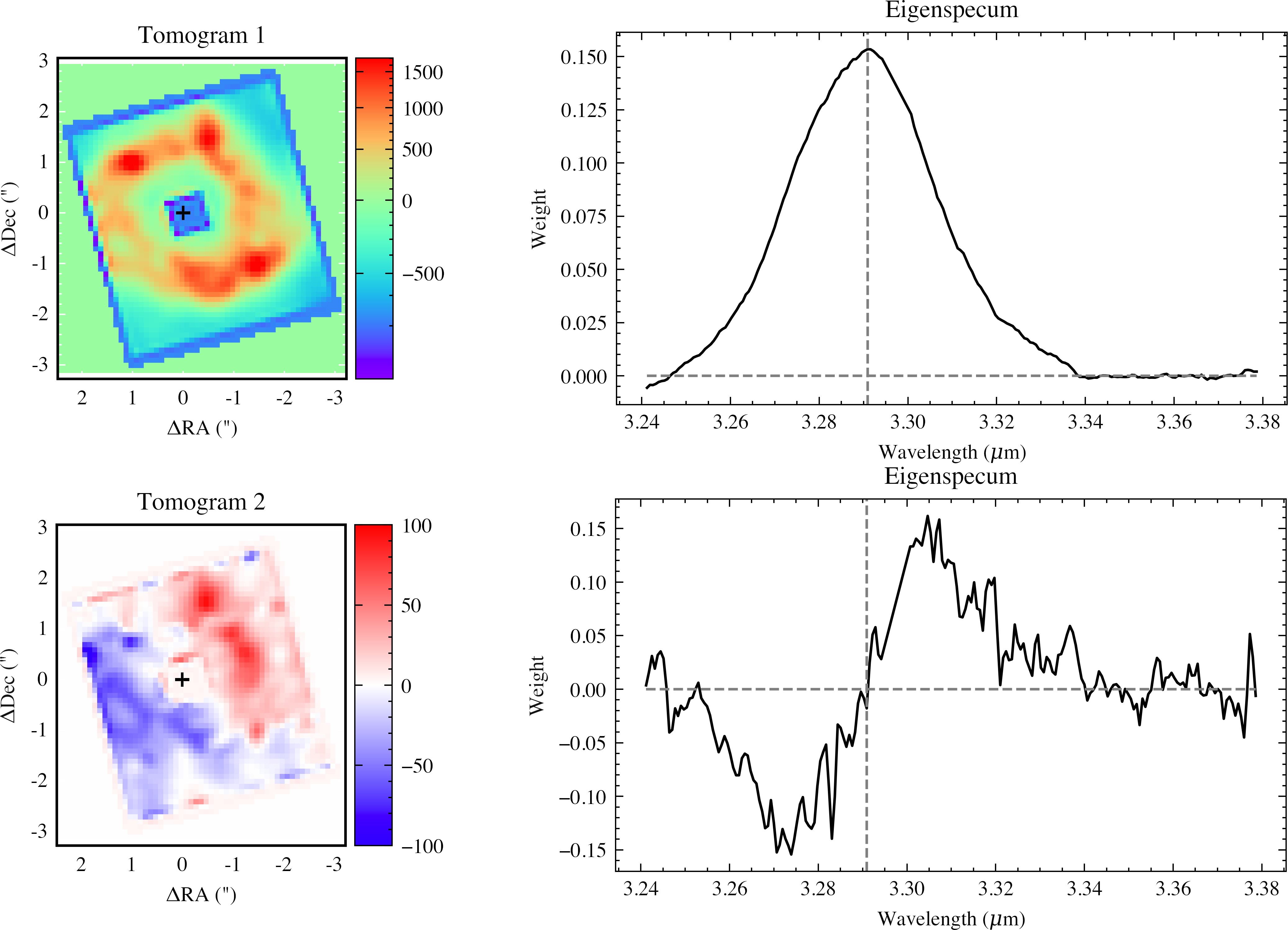}
    \caption{Same as Fig. \ref{fig:NGC3256SBrB} but for the 3.3 $\mu$m PAH in NGC 7469.}
    \label{fig:NGC7469PAH} 
\end{figure*}

In Fig. \ref{fig:NGC3256NH2}, \ref{fig:NGC3256SBrB}, \ref{fig:NGC3256SH2}, \ref{fig:NGC3256SPAH}, \ref{fig:NGC7469BrB}, \ref{fig:NGC7469H2}, \ref{fig:NGC7469PAH} we show the PCA decomposition for the other emission features in this work. 

\bsp	
\label{lastpage}